# Improved mathematical models of structured-light modulation analysis technique for contaminant and defect detection

## YIYANG HUANG, HUIMIN YUE,* YUYAO FANG, YIPING SONG AND YONG LIU

*State Key Laboratory of Electronic Thin Films and Integrated Devices, School of Optoelectronic Science and Engineering, University of Electronic Science and Technology of China, Chengdu, 610054, P. R. China*
*\* yuehuimin@uestc.edu.cn*

**Abstract:** Surface quality inspection of optical components is critical in optical and electronic industries. Structured-Light Modulation Analysis Technique (SMAT) is a novel method recently proposed for the contaminant and defect detection of specular surfaces and transparent objects, and this approach was verified to be effective in eliminating ambient light. The mechanisms and mathematical models of SMAT were analyzed and established based on the theory of photometry and the optical characteristics of contaminants and defects. However, there are still some phenomena exist as conundrums in actual detection process, which cannot be well explained. In order to better analyze the phenomena in practical circumstances, improved mathematical models of SMAT are constructed based on the surface topography of contaminants and defects in this paper. These mathematical models can be used as tools for analyzing various contaminants and defects in different systems, and provide effective instruction for subsequent work. Simulations and experiments on the modulation and the luminous flux of fringe patterns have been implemented to verify the validity of these mathematical models. In adddition, by using the fringe patterns with mutually perpendicular sinusoidal directions, two obtained modulation images can be merged to solve the incomplete information acquisition issue caused by the differentiated response of modulation.

## 1. Introduction

Quality control of optical components has become one of the most important subjects in optical manufacturing industries. Contaminants and defects on the surface of optical components can greatly affect the performance of the device equipped with these components [1]. Therefore, many inspection methods have been proposed for the detection of optical components. These methods can be roughly divided into three categories: microscopic detection method, manual detection method and machine vision method.

Atomic Force Microscope (AFM) and Scanning Electron Microscope (SEM) are two typical microscopic instruments for surface inspection [2, 3]. Despite the high detection accuracy, long time consumption makes them difficult to meet the requirements of industrial inspection.

Considering the speed and accuracy, the detection for optical components in industrial environments is still mainly dominated by the human visual inspection method [4]. However, the efficiency of this method is too low to meet the desired detection requirements.

Machine vision methods usually need to cooperate with different optical principles in practical applications. Polarization characteristics and the single sided diffraction of defects can be exploited to detect defects on optical components [5, 6]. Dark-field imaging is a method that uses the scattering properties of defects to capture the defect information [7]. For these three approaches, if we expect to get more precise detection results, it is necessary to introduce a microscope, which brings about an increase in cost and time.

In recent years, detection methods adopting structured light illumination are rapidly developing [8-11]. The images captured by the camera can be calculated by different algorithms to get corresponding detection results [12-15].

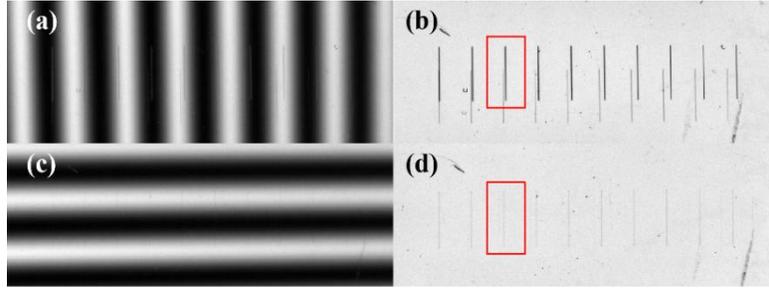

Fig. 1. In the reflection system, (a) one frame of captured fringe patterns and (b) the modulation result of linear defect 1. (c) One frame of captured fringe patterns and (d) the modulation result of linear defect 2.

Structured-light Modulation Analysis Technique (SMAT) is a new method recently proposed for the detection of specular surfaces and transparent objects [16]. In Ref. [16], the mechanisms and mathematical models of SMAT were analyzed and established based on the theory of photometry and the optical characteristics of contaminants and defects. However, in actual detection process, we found that there are some phenomena that can not be quantitatively explained by the mathematical models in Ref. [16], one of which typically is the differentiated response of fringe patterns and modulation image. It can be discovered that when the spatial relation between the defect and the fringe pattern is diverse, the defect on the captured fringe pattern and the modulation image will both be different, as shown in Fig. 1.

To explicate the diversity of the modulation and captured fringe patterns, the mathematical models of SMAT are reconstructed based on the topography of contaminants and defects in this paper. Taking the light reflection system as an example, the difference between the mathematical models in Ref. [16] and the improved mathematical models in this paper is shown as Fig. 2.

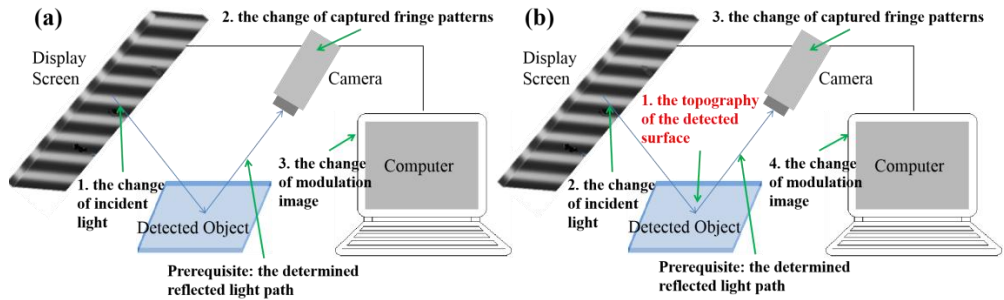

Fig. 2. In the reflection system, the derivation process of (a) the original mathematical models in Ref. [16], (b) the improved mathematical models in this paper.

The derivation process of the mathematical models in Ref. [16] is shown in Fig. 2(a). When the reflected light path is determined, the presence of contaminants and defects will lead to a transformation in incident light. In Ref. [16], based on the change of incident light, the altered reflected fringe patterns can be obtained by using the camera, and then the modulation image can be extracted based on the captured fringe patterns and the modulation equation. The mathematical models calculated according to this process are suitable for analyzing the impact of the change in incident light on the detected results.

In contrast to Ref. [16], in this paper, the biggest variable introduced in the modeling process is the topography of contaminants and defects, and the mathematical models' derivation process is shown in Fig. 2(b). When the reflected light path is determined, the form of incident light will be controlled by the concrete topography of the detected object. According to the change of incident light caused by the change of object topography, the altered reflected fringe patterns can be obtained by the camera, and then the modulation image can be extracted by using the modulation equation and the captured fringe patterns. The mathematical models calculated based on this process are applicable for analyzing the detected results of diverse defects on various objects. Furthermore, they can also be used to analyze the gradient values of different inclined planes, which will not be discussed in detail here.

The mathematical models in Ref. [16] and the mathematical models in this paper have their own focuses, which results in the difference in their applications. For better explain the diversity of the response of contaminants and defects in modulation and captured fringe patterns, it is necessary to consider the topography of contaminants and defects, so the derivation below is performed according to the process shown in Fig. 2(b). In Section 2, the mathematical models of SMAT are established based on the topography of contaminants and defects. In Section 3, the simulated luminous flux and modulation of linear defects show the same differentiation as the actual results shown in Fig. 1, and this difference can also be verified by the experiments in Section 4. In addition, in Section 4, a merged algorithim is proposed to solve the incomplete information acquisition issue caused by the differentiated response of modulation. Section 5 outlines the conclusions of this paper.

## 2. Principle

The basic principle of SMAT is expounded in Section 2.1. The mathematical models for specular surfaces and transparent objects are established in Sections 2.2 and 2.3, respectively.

### 2.1 Mechanism of structured-light modulation analysis technique

The feasibility of SMAT has been attested by the reflection system and the transmission system mentioned in Ref. [16], and the system structures are shown in Fig. 3.

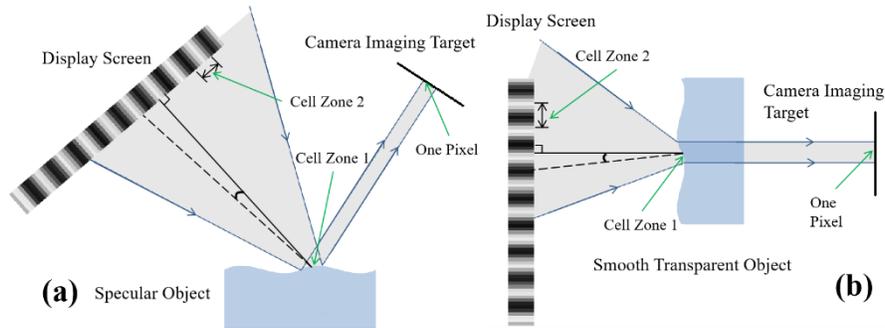

Fig. 3. The system structures of (a) the reflection system, and (b) the transmission system. These figures also show the light relation when there are contaminants and defects on the detected object.

The fringe patterns projected by the display screen can be expressed as:
$$I_n(x, y) = A(x, y) + B(x, y) \cdot \cos[\varphi(x, y) + \delta_n] \tag{1}$$

$A(x, y)$ is the average light intensity of fringe pattern. $B(x, y)$ reflects the contrast of fringe pattern. $\varphi(x, y) = 2\pi f x$ is initial phase distribution. $f$ is spatial frequency. $\delta_n = \frac{n}{N} 2\pi$ is phase shift size. $N$ is the number of phase shift steps.

Assume that the reflectivity of the specular surface and the transmissivity of the transparent object is 100%, the light intensity of the distorted fringe patterns captured by the camera can be expressed as:

$$I_n^{'}(x, y) = A(x, y) + B(x, y) \cdot \cos[\varphi(x, y) + \varphi'(x, y) + \delta_n] \quad (2)$$

$\varphi'(x, y)$ is the additional phase introduced by the detected object.

Based on Eq. (2), the modulation equation can be obtained [16]:

$$M(x, y) = \sqrt{(\sum_{n=0}^{N-1} I_n^{'}(x, y) \cdot \cos\delta_n)^2 + (\sum_{n=0}^{N-1} I_n^{'}(x, y) \cdot \sin\delta_n)^2} \quad (3)$$

The light relation of the reflection system is shown in Fig. 3(a), and this relation is established based on the determined reflected light path. It is assumed that the reflected light rays received by one pixel of camera imaging target come from a small region (Cell Zone 1) on the detected surface. When the detected specular surface is clean and intact, the incident light in Cell Zone 1 comes from a single constant light source (Cell Zone 2) on the display screen. Cell Zone 2 can be regarded as the pixel on the display screen. When there is contaminant or defect, the specular surface shows partially diffuse characteristics, which diverges the light source region from the original Cell Zone 2 to a larger region, as shown in Fig. 3(a). The light relation of the transmission system is similar to that of the reflection system, as shown in Fig. 3(b).

According to the cases of incident light in Fig. 3 and the surface topography of contaminants and defects, the mathematical models of SMAT can be deduced, and the solving procedure is shown in Fig. 2(b). More details about the derivation process are shown as follows.

## 2.2 Improved mathematical models of the reflection system for the detection of specular surfaces with contaminants and defects

In the reflection system, the incident light rays are still analyzed based on the optical path reversible theorem and one pixel in camera imaging target, as shown in Fig. 3(a). The calculation process refers to Fig. 2(b). Cell Zone 1 is the region on the detected object captured by one pixel of camera imaging target. Cell Zone 2 represents a small single constant light source region (a pixel) on the display screen. According to the law of reflection, the angle between the normal line of the display screen and the horizontal plane, and the angle between the camera's optical axis and the horizontal plane, are set to the same value.

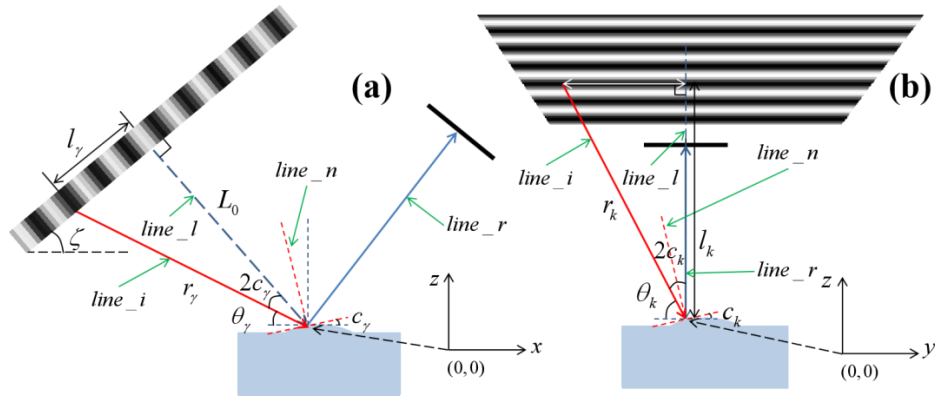

Fig. 4. The relation between incident light rays and reflected light rays (a) on $\gamma$ plane, (b) on $k$ plane.

On the basis of ignoring the influence of Point Spread Function (PSF) of the imaging system, the modulation equation is derived as follows.

The display screen used for illumination is a Lambertian radiator with the characteristic of anisotropic illumination. The illuminance of the planar light source at a distance $r$ is [17]:

$$E = \frac{L dA_s \cos\theta_1 \cos\theta_2}{r^2} \quad (4)$$

$L$ represents the luminance of the planar light source. $\theta_1$ is the solid angle between the normal line of the light source and the light ray emitted by this region. $\theta_2$ is the angle between the normal line of illuminated plane and the light ray from the light source. $dA_s$ is the area of the planar light source. In this paper, let Cell Zone 2 be the planar light source for subsequent calculation, which means that $dA_s$ can be considered as the area of Cell Zone 2, and is set to a constant value $A_s$ here.

Eq. (4) is the core of the derivation process, and the most of the content in Sections 2.2 and 2.3 is basically to calculate the parameters in Eq. (4).

$T$ is the spatial period of the fringe pattern on the display screen. $cell$ indicates the length and width of Cell Zone 2, and there is a relation $A_s = (cell)^2$. $T$ is measured in units of $cell$. $L_0$ represents the distance between the specular surface and the display screen. The angle between the display screen and the horizontal plane is $\zeta$.

Divide Cell Zone 1 into some smaller regions, and these smaller regions can be called Elements. Assume that each of the Elements which reflect the light from the display screen reflects the light emitted by a Cell Zone 2. The ray relations of one Element on $\gamma$ plane and $k$ plane are shown in Fig. 4. On $\gamma$ plane, the inclination angle of the Element compared to the original plane is $c_\gamma$, and that on $k$ plane is $c_k$. The calculation process of the ray relation in this Element is shown as follows.

In Fig. 4, $line\_i$ is the incident light path. $r_\gamma$ and $r_k$ are the projections of $line\_i$ on $\gamma$ and $k$ planes. $\theta_\gamma$ and $\theta_k$ are the projections of the included angle between horizontal plane and $line\_i$ on $\gamma$ and $k$ planes, respectively. From Fig. 4(a), it can be calculated that:

$$r_\gamma = \frac{L_0}{\cos 2c_\gamma} \quad (5)$$

$$\theta_\gamma = \frac{\pi}{2} - \zeta - 2c_\gamma \quad (6)$$

By observing Fig. 4(b), it can be found that:

$$r_k = \frac{l_k}{\cos 2c_k} = \frac{r_\gamma \sin\theta_\gamma}{\cos 2c_k} \quad (7)$$

$$\theta_k = \frac{\pi}{2} - 2c_k \quad (8)$$

$line\_l$ is the normal line of the light-emitting plane. On $k$ plane, $l_k$ is the projection of $line\_i$ on $line\_l$.

Based on Eq. (5), Eq. (7) and Eq. (8), the distance between the light source and the detected surface can be deduced:

$$r^{(R)} = \sqrt{r_\gamma^2 + r_k^2 \cos^2 \theta_k} \tag{9}$$

Combining Eq. (6) and Eq. (8), $line\_i$ can be attained:

$$z = -x\cot(\zeta + 2c_\gamma) = -y\cot 2c_k \tag{10}$$

$line\_l$ is:

$$z = -x\cot\zeta \tag{11}$$

According to the vector equation in solid geometry, the cosine of the angle between $line\_i$ and $line\_l$ is:

$$\cos\theta_1^{(R)} = \frac{\tan(\zeta + 2c_\gamma) + \cot\zeta}{\sqrt{\left[\tan^2(\zeta + 2c_\gamma) + \tan^2 2c_k + 1\right](\cot^2\zeta + 1)}} \tag{12}$$

Based on the geometric relation of light rays in Fig. 4, the normal line of the illuminated plane ($line\_n$) can be obtained:

$$z = -x\cot c_\gamma = -y\cot c_k \tag{13}$$

$\theta_2^{(R)}$ is the angle between $line\_i$ and $line\_n$. According to the law of reflection, the angle between $line\_i$ and $line\_n$ is equal to the angle between $line\_n$ and the reflected light ray ($line\_r$). To simplify the calculation process of $\cos\theta_2^{(R)}$, we can compute the cosine of the angle between $line\_n$ and $line\_r$.

$line\_r$ is:

$$z = x\cot\zeta \tag{14}$$

The cosine of the angle between $line\_r$ and $line\_n$ is:

$$\cos\theta_2^{(R)} = \frac{\cot\zeta - \tan c_\gamma}{\sqrt{\tan^2 c_\gamma + \tan^2 c_k + 1}\sqrt{\cot^2\zeta + 1}} \tag{15}$$

Besides, the phase shift size caused by the object surface is:

$$\varphi^{(R)}(x, y) = 2\pi \frac{l_\gamma}{T \cdot cell} = -2\pi \frac{L_0 \tan 2c_\gamma}{T \cdot cell} \tag{16}$$

$l_\gamma$ is the distance of light source change caused by the object surface on $\gamma$ plane.

The luminance of the planar light source (Cell Zone 2) is:

$$L_n^{(R)} = A(x, y) + B(x, y)\cos\left[\varphi(x, y) + \delta_n + \varphi^{(R)}(x, y)\right] \tag{17}$$

In Cell Zone 1, some Elements reflect the light come from the display screen, and others reflect the light from environment. It is assumed that each of the Elements which reflect the light from the display screen reflects the light emitted by a Cell Zone 2. Let the illuminance from environment be $C(x, y)$. Based on Eq. (4) and above results, the luminous flux captured by one pixel of imaging target can be obtained:

$$\Phi_{nR}^{(R)} = \alpha\left(\sum_{k=c_{k0}}^{c_k}\sum_{\gamma=c_{\gamma0}}^{c_\gamma} s_{\gamma k} A_s \frac{L_n^{(R)}}{\left(r^{(R)}\right)^2}\cos\theta_1^{(R)}\cos\theta_2^{(R)} + \sum_{k=c_{k0}^*}^{c_k^*}\sum_{\gamma=c_{\gamma0}^*}^{c_\gamma^*} s_{\gamma k}^* C(x, y)\right) \tag{18}$$

$\alpha$ is the surface reflectivity of the specular surface. $s_{\gamma k}$ indicates the area of Elements which reflect light from the display screen, $s_{\gamma k}^*$ represents the area of Elements which reflect light from environment.

The restrictions for light come from the display screen are:

$$S_{\gamma 1} < -L_0 \tan 2c_\gamma < S_{\gamma 2} \tag{19}$$

$$S_{k1} < -l_k \tan 2c_k = -r_\gamma \sin\theta_\gamma \tan 2c_k < S_{k2} \tag{20}$$

$S_{\gamma 1}$ and $S_{\gamma 2}$ are the negative and positive range of the display screen compared to the original light source position on $\gamma$ plane. $S_{k1}$ and $S_{k2}$ are the negative and positive range of the display screen compared to the original light source position on $k$ plane.

If the specular surface is clean and intact, the area of Cell Zone 1 is $s_0$, and the surface reflectivity of Cell Zone 1 is $\alpha_0$, the reflected luminous flux Eq. (18) changes to:

$$\Phi_{n0}^{(R)} = \frac{\alpha_0 s_0 A_s \cos\zeta}{L_0^2}\{A(x,y) + B(x,y)\cos[\varphi(x,y) + \delta_n]\} \tag{21}$$

By substituting Eq. (18) into Eq. (3), the photometric modulation equation can be obtained:

$$M^{(R)}(x,y) = \frac{\alpha A_s N B(x,y)}{2} \sqrt{\left(\sum_{k=0}^{c_k}\sum_{\gamma=0}^{c_\gamma} s_{\gamma k} \frac{\cos\theta_1^{(R)}\cos\theta_2^{(R)}}{\left(r^{(R)}\right)^2}\cdot\cos\varphi_m^{(R)}\right)^2 + \left(\sum_{k=0}^{c_k}\sum_{\gamma=0}^{c_\gamma} s_{\gamma k} \frac{\cos\theta_1^{(R)}\cos\theta_2^{(R)}}{\left(r^{(R)}\right)^2}\cdot\sin\varphi_m^{(R)}\right)^2} \tag{22}$$

$$\varphi_m^{(R)} = \varphi(x,y) + \varphi^{(R)}(x,y) \tag{23}$$

If the specular surface is clean and intact, Eq. (22) can be simplified as:

$$M_0^{(R)}(x,y) = \frac{\alpha_0 s_0 A_s N B(x,y)\cos\zeta}{2L_0^2} \tag{24}$$

## 2.3 Improved mathematical models of the transmission system for the detection of transparent objects with contaminants and defects

For transparent objects whose front and back surfaces are parallel to each other and only one surface has contaminants or defects, the relation of light rays is shown in Fig. 3(b). The calculation process is nearly the same as Fig. 2(b). Set the plane of the object parallel to the screen plane, and make the optical axis of camera is perpendicular to the object plane and the screen plane.

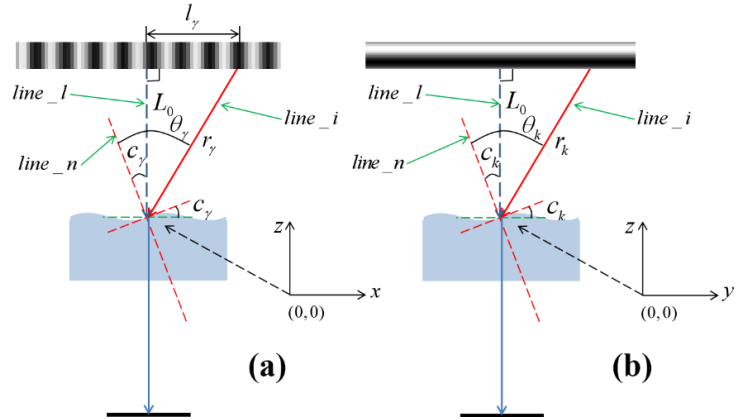

Fig. 5. The relation between incident light rays and transmitted light rays (a) on $\gamma$ plane, (b) on $k$ plane.

On the basis of ignoring the influence of Point Spread Function (PSF) of the imaging system, the modulation equation is derived as follows.

The incident light rays and transmitted light rays satisfy the law of refraction:

$$\frac{\sin\theta}{\sin c} = n \tag{25}$$

$c$ is the angle between the normal line of the interface of mediums and the light ray in medium 2. $\theta$ indicates the angle between the normal line of the interface of mediums and the light ray in medium 1. $n$ is the relative refraction index of medium 2 to medium 1.

$T$ is the spatial period of the fringe pattern, $cell$ indicates the length and width of Cell Zone 2. $L_0$ represents the distance between the detected transparent object and the display screen. $A_s$ is the area of Cell Zone 2.

Divide Cell Zone 1 into some smaller regions (Elements). The inclination angle of the Element on $\gamma$ plane is $c_\gamma$, and that on $k$ plane is $c_k$. The calculation process of the ray relation in one Element is shown as follows.

In Fig. 5, $line\_i$ is the incident light path, and $line\_n$ is the normal line of the illuminated plane. $r_\gamma$ and $r_k$ are the projections of $line\_i$ on $\gamma$ and $k$ planes. $\theta_\gamma$ and $\theta_k$ are the projections of the included angle between $line\_i$ and $line\_n$ on $\gamma$ and $k$ planes, respectively. By observing Fig. 5(a), it can be seen that:

$$\theta_\gamma = \arcsin(n\sin c_\gamma) \tag{26}$$

From Fig. 5(b), it can be calculated that:

$$\theta_k = \arcsin(n\sin c_k) \tag{27}$$

Combining Eq. (26) and Eq. (27), $line\_i$ can be obtained:

$$z = x\cot(\theta_\gamma - c_\gamma) = y\cot(\theta_k - c_k) \tag{28}$$

$line\_n$ is:

$$z = -x\cot c_\gamma = -y\cot c_k \tag{29}$$

The normal line of the light-emitting surface ($line\_l$) is:

$$x = y = 0 \tag{30}$$

The cosine of the included angle between $line\_i$ and $line\_n$ can be obtained:

$$\cos\theta_2^{(T)} = \frac{e_1 e_3 + e_2 e_4 + 1}{\sqrt{(e_1^2 + e_2^2 + 1)(e_3^2 + e_4^2 + 1)}} \tag{31}$$

$$e_1 = \tan(\theta_\gamma - c_\gamma) \tag{32}$$

$$e_2 = \tan(\theta_k - c_k) \tag{33}$$

$$e_3 = -\tan c_\gamma \tag{34}$$

$$e_4 = -\tan c_k \tag{35}$$

The cosine of the included angle between $line\_i$ and $line\_l$ is:

$$\cos\theta_1^{(T)} = \frac{1}{\sqrt{e_1^2 + e_2^2 + 1}} \tag{36}$$

The distance between the light source and the detected object can be calculated:

$$r^{(T)} = L_0\sqrt{e_1^2 + e_2^2 + 1} \tag{37}$$

The luminance of the planar light source (Cell Zone 2) becomes:

$$L_n^{(T)} = A(x,y) + B(x,y)\left[\varphi(x,y) + \delta_n + \varphi^{(T)}(x,y)\right] \tag{38}$$

$$\varphi^{(T)}(x,y) = \frac{2\pi l_\gamma}{T \cdot cell} = \frac{2\pi L_0 e_1}{T \cdot cell} \tag{39}$$

$l_\gamma$ is the distance of light source change caused by the object surface on $\gamma$ plane.

Let the illuminance from environment be $C(x,y)$. Based on the relation who is similar to the relation between Cell Zone 1 and Elements in the reflection system, the luminous flux captured by one pixel of camera imaging target can be deduced according to Eq. (4) and above results:

$$\Phi_{nT}^{(T)} = \beta\left(\sum_{k=0}^{c_k}\sum_{\gamma=0}^{c_\gamma} s_{\gamma k} A_s \frac{L_n^{(T)}}{\left(r^{(T)}\right)^2}\cos\theta_1^{(T)}\cos\theta_2^{(T)} + \sum_{k=0}^{c_k^*}\sum_{\gamma=0}^{c_\gamma^*} s_{\gamma k}^* C(x,y)\right) \tag{40}$$

$\beta$ is the transmissivity of the transparent object. $s_{\gamma k}$ indicates the area of Elements which transmit light from the display screen, $s_{\gamma k}^*$ represents the area of Elements which transmit light from environment.

The restrictions for light come from the display screen are:

$$S_{\gamma 1} < L_0 e_1 < S_{\gamma 2} \tag{41}$$

$$S_{k1} < L_0 e_2 < S_{k2} \tag{42}$$

$S_{\gamma 1}$ and $S_{\gamma 2}$ are the negative and positive range of the display screen compared to the original light source position on $\gamma$ plane. $S_{k1}$ and $S_{k2}$ are the negative and positive range of the display screen compared to the original light source position on $k$ plane.

In addition, there is an additional limitation of the total reflection in the transmission system. When the inclination angle between the changed surface and the original surface is greater than a critical angle, the camera will receive the reflected light from environment instead of the transmitted light from the display screen. The critical angle is:

$$c_T = \arcsin\left(\frac{1}{n}\right) \tag{43}$$

If the transparent object is clean and intact, the area of Cell Zone 1 is $s_0$, the transmissivity of Cell Zone 1 is $\beta_0$, the luminous flux Eq. (40) changes to:

$$\Phi_{n0}^{(T)} = \frac{\beta_0 s_0 A_s}{L_0^2}\left\{A(x,y) + B(x,y)\cos\left[\varphi(x,y) + \delta_n\right]\right\} \tag{44}$$

By substituting Eq. (40) into Eq. (3), the photometric modulation equation can be obtained:

$$M^{(T)}(x,y) = \frac{\beta A_s NB(x,y)}{2}\sqrt{\left(\sum_{k=0}^{c_k}\sum_{\gamma=0}^{c_\gamma} s_{\gamma k}\frac{\cos\theta_1^{(T)}\cos\theta_2^{(T)}}{\left(r^{(T)}\right)^2}\cdot\cos\varphi_m^{(T)}\right)^2 + \left(\sum_{k=0}^{c_k}\sum_{\gamma=0}^{c_\gamma} s_{\gamma k}\frac{\cos\theta_1^{(T)}\cos\theta_2^{(T)}}{\left(r^{(T)}\right)^2}\cdot\sin\varphi_m^{(T)}\right)^2} \tag{45}$$

$$\varphi_m^{(T)} = \varphi(x,y) + \varphi^{(T)}(x,y) \tag{46}$$

If the transparent object is clean and intact, Eq. (45) can be simplified as:

$$M_0^{(T)}(x, y) = \frac{\beta_0 s_0 A_s NB(x, y)}{2L_0^2} \tag{47}$$

## 3. Simulations

In order to verify the correctness of above derivation and facilitate the verification of simulation results, the following simulations are carried out from two aspects: the luminous flux of the single-frame captured fringe image and the final modulation image. To observe the differences of different contaminants and defects, the simulations are performed based on different linear defects below. The topography of defects refers to Ref. [18].

The region on the detected object captured by one pixel of camera imaging target (Cell Zone 1) can be divided into 100*100 smaller regions (Elements). The surface topography of defects is shown in Fig. 6.

The linear defect 1 can be expressed as:

$$T_1 = 100 - w\sin\frac{\pi x}{100} \tag{48}$$

The linear defect 2 can be expressed as:

$$T_2 = 100 - w\sin\frac{\pi y}{100} \tag{49}$$

In Eq. (48) and Eq. (49), $w$ is the parameter that controls the overall curvature of Cell Zone 1.

The fringe structured light is:

$$I_n(x, y) = A + B \cdot \cos(2\pi fx + \delta_n) \tag{50}$$

Let $\varphi = 2\pi fx$. $\varphi$ is the initial phase of fringe patterns. The relations between different linear defects and the sinusoidal direction of fringe structured light are shown in Fig. 7(a), and the different initial phase points on the fringe pattern are shown in Fig. 7(b).

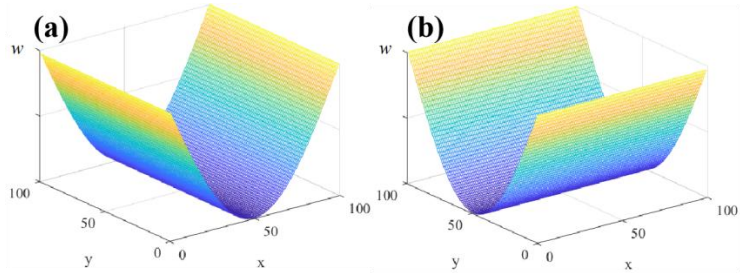

Fig. 6. The surface topography of (a) linear defect 1, (b) linear defect 2.

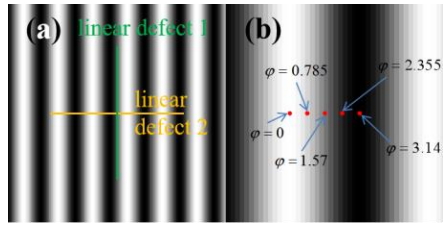

Fig. 7. (a) Spatial relation between linear defects and fringe structured light. (b) The positions of different initial phase points in fringe structured light.

The calculation process of luminous flux and modulation is:
(1) Find the radians of different Elements in $x$ and $y$ directions.

(2) The luminous flux values of the defect can be obtained by using Eq. (18) or Eq. (40), and the modulation results can be calculated by Eq. (22) or Eq. (45).

(3) Calculate the original luminous flux value from Eq. (21) or Eq. (44), and the original modulation value from Eq. (24) or Eq. (47).

(4) The relative luminous flux results can be obtained by subtracting the original luminous flux value obtained in step (3) from the luminous flux results obtained in step (2) and multiplying the attained differences by a constant. The relative luminous flux values are normalized to the range of -1 to 1. If the luminous flux value obtained in step (2) is equal to the original luminous flux value in step (3), the relative luminous flux value is 0.

(5) The modulation results in step (2) can be divided by the original modulation value in step (3) to obtain the relative modulation results. The relative modulation values are normalized to the range of 0 to 1. If the modulation value in step (2) is equivalent to the original modulation value in step (3), the relative modulation value is 1.

The luminous flux captured by one pixel of the fringe pattern under different conditions can be calculated by adjusting $w$ and $\varphi$. Meantime, the influence of different initial phases on modulation results can be ignored, thus the modulation in different situations can be obtained by only altering $w$.

In the simulations described below, the changes in reflectivity and transmissivity caused by contaminants and defects are temporarily ignored. The detected material simulated here is a glass plate with $n=1.5$.

### 3.1 Luminous flux of the captured fringe image

The luminous flux results of linear defect 1 and linear defect 2 are shown in Fig. 8. Since there are only slight differences between the reflection system and the transmission system, their simulation results are concluded together.

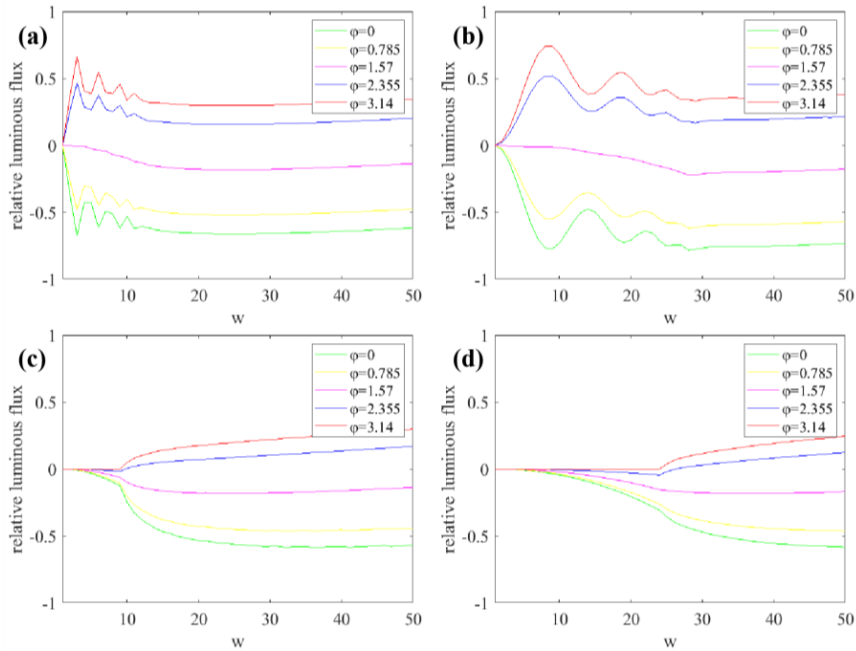

Fig. 8. The relative luminous flux of linear defect 1 in (a) the reflection system, (b) the transmission system. The relative luminous flux of linear defect 2 in (c) the reflection system, (d) the transmission system.

From the single line in Fig. 8, it can be found that with the increase of overall curvature $w$, the relative luminous flux of linear defects all will change, and the change direction is related to the initial phase $\varphi$. The luminous flux results of linear defect 1 alter abruptly at first, while the results of linear defect 2 change slowly. With the continuous increase of $w$, the luminous flux results of linear defect 1 alter tardily, yet those of linear defect 2 vary faster than before.

By observing the different lines in Fig. 8, it can be discovered that the luminous flux value will be lower than the original value when the defect is at the bright place of structured light, while it will be higher than the original value when the defect is at the dark place. When the defect is at the place with moderate luminance, the luminous flux value of the defect is almost the same as the original value.

Compared with the luminous flux results of linear defect 1, linear defect 2 causes weaker changes in luminous flux results, which means that linear defect 2 is less observable than linear defect 1 on the captured fringe image, and this phenomenon is called the differentiated response of fringe patterns.

### 3.2 Comparison of modulation

By observing the modulation results, it is discovered that there are some differences between different defects in the same system. For the sake of comparison, the modulation results are embodied as Fig. 9.

From Fig. 9, it can be discovered that the modulation value of linear defect 1 is obviously lower than that of linear defect 2 when $w$ is the same. Lower value equals to higher contrast between defects and intact places, which means that the modulation result of linear defect 1 is more obvious than that of linear defect 2, and this phenomenon is called the differentiated response of modulation.

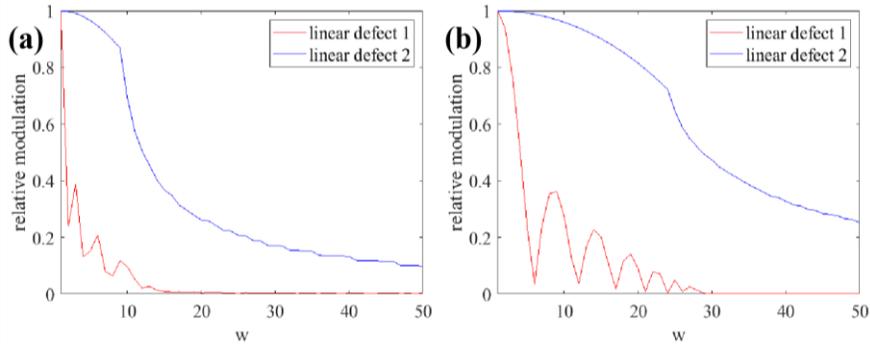

Fig. 9. Modulation comparison of linear defect 1 and linear defect 2 in (a) the reflection system, (b) the transmission system.

## 4. Experimental work

The modulation values of the sites with contaminants or defects are lower than that of the clean and intact places, which has been verified in Ref. [16] and will not be repeatedly discussed here. In this section, the experiments are conducted for observing the differentiated response of modulation and captured fringe patterns. The systems used for experiments are shown in Fig. 3.

The reflection system and the transmission system both comprise a display screen (Philips 246V6QSB, 1920×1080 pixels, 0.272 $mm$ pixel spacing in both directions) and a single camera (AVT Manta G-917B, 25 $mm$ focal length, 3384×2710 pixels, 5.5 $\mu m$ pixel spacing). A glass plate is used as the detected sample in following experiments. There are ten scratches on the glass plate, which were made by chemical sketching. The width of the scratches on the glass plate varies from 5 to 50 $\mu m$, with the length 5 $mm$ and the depth 2 $\mu m$. The difference in width between two adjacent scratches is 5 $\mu m$.

*4.1 The differentiated responses of modulation and captured fringe patterns*

Two series of fringe patterns whose sinusoidal direction is perpendicular to each other are projected for image capture, and the captured fringe patterns and calculated modulation results are shown in Fig. 1 and Fig. 10. The grayscale of these Figs. is stretched to 0-255.

From Figs. 1(a) and 10(a), it can be discovered that when the scratches are the type of linear defect 1, the scratches are visible in the captured fringe image, and the scratches match the law of darkening bright places and brightening dark places. In addition, in the case when the scratches are the type of linear defect 2, the scratches are barely observable in the captured fringe pattern, as shown in Figs. 1(c) and 10(c). These results are consistent with the simulation results in Fig. 8.

By comparing Figs. 1(b), 1(d), 10(b) and 10(d), it can be seen that the modulation results of linear defect 1 are more obvious than those of linear defect 2, which is also the same as the simulation results in Fig. 9. The lighter scratches in Fig. 1(b) are the second image caused by parasitic fringes [19], which will not be detailed discussed here.

In a few words, from the experimental results in Figs. 1 and 10, it can be concluded that linear defect 1 is more observable than linear defect 2 in both captured fringe patterns and modulation image, which is consistent with the simulation results in Figs. 8 and 9.

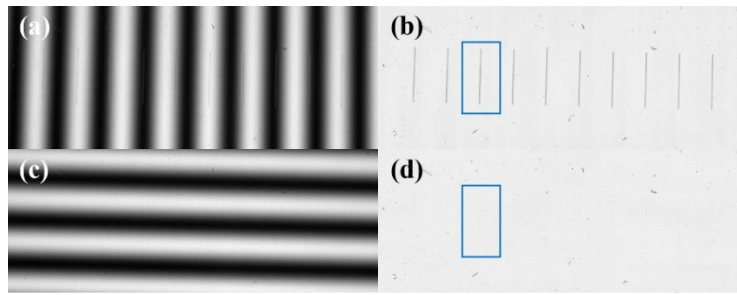

Fig. 10. In the transmission system, (a) one frame of captured fringe patterns and (b) the modulation result of linear defect 1. (c) One frame of captured fringe patterns and (d) the modulation result of linear defect 2.

*4.2 The merged algorithm*

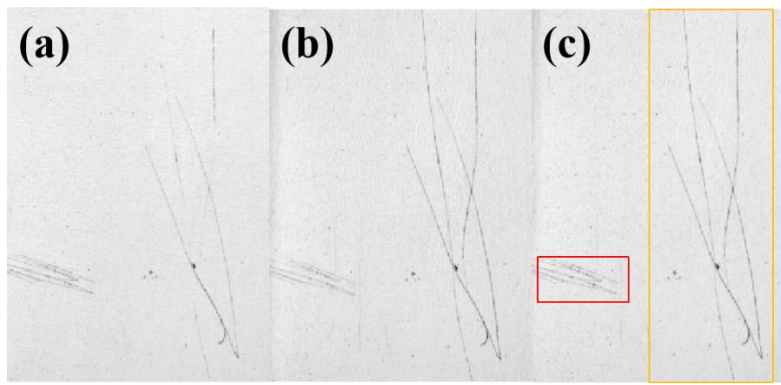

Fig. 11. The modulation result obtained (a) in the first direction, (b) in the second direction, (c) after merging modulation results. In (c), the scratches in the red box mainly come from (a), while the scratches in the orange box principally use the information of (b).

By observing the modulation results in Figs. 1 and 10, it can be found that in some cases, there will be some defects that can not be collected by the modualtion image. So, in order to detect defects at different situations, two series of fringe patterns with mutually perpendicular sinusoidal directions are projected for obtaining different modulation images, and these obtained modulation images can be simply added together to obtain a brand-new image. The advantages of different modulation images are extracted and merged into this new modulation image to solve the incomplete information acquisition issue in original modulation results. Figs. 11(a)-11(b) show the modulation results based on fringe patterns with mutually perpendicular sinusoidal directions, and the merged result is shown in Fig. 11(c). Fig. 11(c) shows better detection effect than Figs. 11(a) and 11(b).

## 5. Conclusion

In this study, the improved mathematical models of luminous flux and modulation in one pixel of camera imaging target are constructed based on the theory of photometry and the topography of contaminants and defects. These mathematical models can be used as analytical tools for SMAT in practical applications. Simulations and experiments on the modulation and the luminous flux of fringe patterns have been conducted to verify the validity of these mathematical models. Besides, in order to detect defects at different situations, in this paper, a merged algorithm is proposed to solve the incomplete information acquisition issue caused by the differentiated response of modulation.

**Funding**

This work was funded by National Nature Science Foundation of China (No. 61875035), Applied Basic Research Program of Sichuan Province (2018JY0579).

**Disclosures:** The authors declare no conflicts of interest.

**References**

1. B. Ma, Y. Zhang, H. Ma, H. Jiao, X. Cheng, and Z. Wang, "Influence of incidence angle and polarization state on the damage site characteristics of fused silica," Appl. Opt. **53**, A96-A102 (2014).
2. S. Gomez, K. Hale, J. Burrows and B. Griffiths, "Measurements of surface defects on optical components," Meas. Sci. Technol. **9**(4), 607–616 (1998).
3. H. Ota, M. Hachiya, Y. Ichiyasu, and T. Kurenuma, "Scanning surface inspection system with defect-review SEM and analysis system solutions," Hatachi Review **55**, 78-82 (2006).
4. X. Tao, Z. Zhang, F. Zhang, and D. Xu, "A novel and effective surface flaw inspection instrument for large-aperture optical elements," IEEE Trans. Instrum. Meas. **64**, 2530-2540 (2015).
5. T. A. Germer, "Angular dependence and polarization of out-of-plane optical scattering from particulate contamination, subsurface defects, and surface microroughness," Appl. Opt. **36**, 8798-8805 (1997).
6. V. M. Schneider, M. Meljnek, and K. T. Gahagan, "Fast detection of single sided diffracted defects in display glass," Measurement, **42**(4), 638-644 (2009).
7. D. Liu, S. Wang, P. Cao, L. Li, Z. Cheng, X. Gao, and Y. Yang, "Dark-field microscopic image stitching method for surface defects evaluation of large fine optics," Opt. Express **21**, 5974-5987 (2013).
8. Q. Zhang, X. Su, L. Xiang, X. Sun, "3-D shape measurement based on complementary Gray-code light", Opt Laser Eng. **50**(4), 574-579 (2012).
9. S. S. Martínez, J. G. Ortega, J. G. García, A. S. García, "A machine vision system for defect characterization on transparent parts with non-plane surfaces," Mach. Vision Appl. **23**(1), 1-13 (2012).
10. H. Yue, H. G. Dantanarayana, Y. Wu, J. M. Huntley, "Reduction of systematic errors in structured light metrology at discontinuities in surface reflectivity," Opt Laser Eng. **112**, 68-76 (2019).
11. K. Xue, Y. Li, S. Lu, and L. Chen, "Three-dimensional shape measurement using improved binary spatio-temporal encoded illumination and voting algorithm," Appl. Opt. **50**, 5508-5512 (2011).
12. K. Qian, "Two-dimensional windowed Fourier transform for fringe pattern analysis: Principles, applications and implementations," Opt Laser Eng. **45**(2), 304-317 (2007).
13. P. Zhao, N. Gao, Z. Zhang, F. Gao, X. Jiang. "Performance analysis and evaluation of direct phase measuring deflectometry," Opt Laser Eng. **103**, 24-33 (2018).

14. Z. Zhang, Y. Wang, S. Huang, Y. Liu, C. Chang, F. Gao, X. Jiang. "Three-dimensional shape measurements of specular objects using phase-measuring deflectometry," Sensors, **17**(12), 2835 (2017).
15. S. Zhang, "High-speed 3D shape measurement with structured light methods: A review", Opt Laser Eng, **106**, 119-131 (2018).
16. Y. Huang, H. Yue, Y. Fang, W. Wang, and Y. Liu, "Structured-light modulation analysis technique for contamination and defect detection of specular surfaces and transparent objects," Opt. Express **27**, 37721-37735 (2019).
17. J. E. Greivenkamp, *Field Guide to Geometrical Optics* (SPIE, 2004).
18. F. Zhang, C. Li, B. Meng, "Investigation of Surface Deformation Characteristic and Removal Mechanism for K9 Glass Based on Varied Cutting-depth Nano-scratch," Chin. J. Mech. Eng. **52**(17), 65-71 (2016). (in Chinese)
19. L. Huang, A. Krishna, "Phase retrieval from reflective fringe patterns of double-sided transparent objects," Meas. Sci. Technol. **23**(8), 085201-85206 (2012).